\documentclass[conference]{IEEEtran}

\AtBeginDocument{%
  \providecommand\BibTeX{{%
    Bib\TeX}}}

\usepackage{graphicx}
\usepackage{textcomp}
\usepackage{multirow}
\usepackage{xcolor}
\usepackage{pgfplots}
\usepackage{booktabs}
\usepackage{subcaption}
\usepackage[caption=false,font=footnotesize]{subfig}
\pgfplotsset{compat=1.18}
\usepackage{soul}
\usepackage{bbm}
\usepackage{amsfonts}
\usepackage{array}
\usepackage{setspace}
\usepackage{url}
\usepackage{amsmath}

\usepackage{tikz}
\usepgfplotslibrary{statistics}

\usepackage{siunitx}
\usepackage{tabularx}

\usepackage[ruled, linesnumbered, vlined, longend]{algorithm2e}
\SetAlFnt{\footnotesize}
\SetAlCapNameFnt{\small}
\SetAlCapFnt{\small}
\IncMargin{3pt}
\SetAlgoNlRelativeSize{-0.5}
\SetKwProg{Def}{def}{$\,$:}{}
\SetKwProg{Func}{Func}{$\,$:}{}
\SetKwInOut{Input}{Input}
\SetKwInOut{Output}{Output}

\def\BibTeX{{\rm B\kern-.05em{\sc i\kern-.025em b}\kern-.08em
    T\kern-.1667em\lower.7ex\hbox{E}\kern-.125emX}}

\begin{document}

\title{CascadeLUT: Information-Ordered Streaming Inference for Bandwidth-Constrained FPGAs}

\author{
Oliver Cassidy, Marta Andronic and George A.~Constantinides\\
Department of Electrical and Electronic Engineering,
Imperial College London, UK\\
Email: \{olly.cassidy23, marta.andronic18, g.constantinides\}@imperial.ac.uk
}

\maketitle

\begin{abstract}

Mapping neural networks to FPGAs enables low-latency, energy-efficient inference, particularly for lookup table (LUT)-based models that eliminate multipliers and map directly to reconfigurable fabric. While prior work achieves high compute efficiency, it typically assumes full-sample availability, causing pipeline stalls in bandwidth-limited streaming scenarios. Here, the bottleneck shifts from computation to data movement, as large input transfers limit throughput and energy efficiency. We present CascadeLUT, an information-structured inference framework organized around bandwidth constraints. Instead of buffering the full input, features are partitioned into ordered subsets and predictions are progressively refined as subsets arrive. The cascade statically controls which layers consume incoming features, enabling deterministic streaming inference without runtime branching. By co-designing feature scheduling with hardware dataflow, CascadeLUT reduces data movement while maintaining accuracy. Across datasets, it achieves 4.0--12.5\(\times\) lower latency, 3.0--5.0\(\times\) higher throughput and up to 13.8\(\times\) lower energy/sample than prior LUT baselines, using 1.2--4.4\(\times\) the LUTs of the smallest DWN baseline per task. We also demonstrate on-device input quantization integrated with LUT-based inference and present end-to-end FPGA results on real-world workloads, with 5\(\times\) reductions in quantization overhead.

\end{abstract}

\begin{IEEEkeywords}
FPGA, LUT-based neural networks, bandwidth-constrained inference, streaming inference, feature scheduling
\end{IEEEkeywords}

\section{Introduction}

Field-programmable gate arrays (FPGAs) are a compelling platform for neural network (NN) inference in latency- and energy-critical settings. By exploiting fine-grained parallelism and mapping arithmetic directly into reconfigurable logic, FPGA designs achieve high throughput at low power~\cite{finn}. This advantage is amplified for LUT-based NNs, where inference is expressed as low-precision operations and implemented directly using LUT resources, eliminating multipliers and simplifying datapaths. In this regime, computation becomes inexpensive relative to data movement~\cite{efficientdnn}, shifting the\,bottleneck.

As a result, these systems are often not compute-bound but input-rate-limited. When inputs arrive over a bandwidth-constrained link---for example, from off-chip memory, sensors or interconnect---the cost of transferring features can exceed that of evaluation.
Conventional FPGA inference pipelines assume full-sample availability, buffering the entire input before execution.
Under bandwidth constraints, this leads to idle compute cycles and increased end-to-end latency.

We present CascadeLUT, a bandwidth-aware inference architecture that co-designs feature ordering, network topology and FPGA dataflow for streaming deployments. Instead of buffering the full input, features are partitioned into ordered subsets and streamed under a fixed bandwidth constraint. The network is organized as a statically scheduled cascade: selected layers consume newly arriving features, while others propagate intermediate activations. Computation begins as soon as the first features arrive, overlapping data movement with inference, so latency scales with load stages rather than input size. Unlike early exit or other adaptive approaches, CascadeLUT introduces no runtime branching; the feature schedule, cascade topology and pipeline are fixed at synthesis time, enabling deterministic latency aligned with input bandwidth.

Across multiple datasets and configurations, CascadeLUT achieves 4.0--12.5\(\times\) latency reductions and 3.0--5.0\(\times\) higher throughput than prior ultra-low-latency FPGA inference architectures while maintaining comparable accuracy. Energy per sample is reduced by up to 13.8\(\times\) on bandwidth-limited workloads due to shorter inference. This uses 1.2--4.4\(\times\) the LUTs of the smallest DWN baseline per task.
Further, we show that preprocessing integrated as LUTs in the fabric adds modest area overhead, kept small by feature truncation. These results show that co-designing feature transport, representation and computation around streaming constraints improves FPGA inference efficiency in both end-to-end latency and energy.

In summary, the key contributions of this paper are:

\begin{itemize}
    \item An open-source toolflow\footnote{https://github.com/ollycassidy13/CascadeLUT} that implements a novel statically scheduled cascade architecture, organizing computation around a learned feature schedule and enabling inference directly on streaming data as it arrives on-chip.

    \item A feature scheduling method that derives a transmission order prioritizing salient features under a fixed bandwidth.
    
    \item We demonstrate that learned preprocessing can be implemented in FPGA fabric with modest area overhead, and report hardware results for an end-to-end inference pipeline that could directly consume sensor data by integrating on-device input quantization with NN execution.

    \item An FPGA implementation on a Zynq Z-7045, evaluated on MNIST, FashionMNIST, KWS and ToyADMOS, achieving 4.0--12.5\(\times\) lower latency and 3.0--5.0\(\times\) higher throughput than prior low-latency inference architectures.
\end{itemize}

\section{Background}\label{sec:background}

FPGA-oriented machine learning frameworks such as FINN~\cite{finn} and hls4ml~\cite{hls4ml,hls4ml-bin} demonstrate that extreme quantization enables highly efficient neural network inference on reconfigurable hardware. By restricting computation to low-precision operations, these systems map naturally onto FPGA logic resources and achieve high throughput and low latency. Subsequent work has further shown that neural network inference can be implemented directly in lookup-table fabric, enabling ultra-low-latency deployments in real-time environments such as high-energy physics experiments at CERN.

Several architectures adopt this LUT-based approach. LogicNets~\cite{logicnets} map individual neurons directly to LUTs, using logical LUTs (L-LUTs) which implement higher-level functions by combining multiple physical LUTs. PolyLUT~\cite{poly, polylut2} absorbs polynomial functions, while NeuraLUT~\cite{NeuraLUT} absorbs small subnetworks into larger L-LUTs to increase representational capacity.
NeuraLUT-Assemble~\cite{assemble}, used as the backbone in this work, constructs neurons as trees of smaller L-LUTs, using learned sparse connectivity to increase effective fan-in while controlling LUT growth. After training, each subnetwork is enumerated into Boolean logic and mapped to a pipelined FPGA implementation.
Related work on weightless neural networks (DWN)~\cite{dwn} builds upon ULEEN~\cite{uleen} to realize inference through collections of learned LUT structures. DiffLogicNet~\cite{difflogic} instead introduces differentiable logic gate networks, enabling gradient-based training of discrete logic models.
Post-training optimization techniques~\cite{reducedlut,compressedlut,yukio}
demonstrate that LUT-based networks can be further compressed to improve area efficiency while preserving accuracy.
Across these architectures, the inference pipeline typically assumes that the complete input vector is available before computation begins. In bandwidth-constrained deployments where off-chip I/O or interconnect limits the rate at which features can be delivered to the compute fabric, this introduces unnecessary idle cycles and increases end-to-end latency. 

\section{CascadeLUT Architecture}

To illustrate the proposed neural network topology, consider the three-stage CascadeLUT architecture in Figure~\ref{fig:lut-diagram}. Input features are streamed to the FPGA through a bandwidth-limited interface that transfers $B_{\mathrm{io}}$ bits per cycle, causing features to arrive in fixed-size subsets. The cascade is built on a NeuraLUT-Assemble~\cite{assemble} backbone.

In the first stage, the most salient subset $S_1$ arrives and is consumed by the first cascade layer, producing hidden activations that propagate through subsequent layers. Weight magnitudes determine feature salience from an initial fully connected training phase (Section~\ref{subsec:feat_order}). In subsequent cycles, subsets $S_2$ and $S_3$ arrive sequentially and are concatenated with intermediate activations to progressively refine the representation, with a final layer producing the output.

Each pipeline stage consumes a feature subset, so computation begins as data arrives and overlaps with feature transfer. This removes full-input buffering, so latency scales with load stages rather than input size. The execution follows a fully static schedule, enabling deterministic feed-forward dataflow without runtime control. All operations map to combinational LUT logic between registers, allowing efficient pipelining and high clock frequency on FPGA fabric.

We assume the feature order is imposed before the bandwidth-limited link, e.g., by storing samples in that order or packetizing sensor-derived features in the learned order. CascadeLUT does not buffer full samples on FPGA for reordering or perform any runtime permutation; the order is fixed across samples and not selected adaptively at runtime. 

\section{Methodology}
\label{sec:method}

In bandwidth-limited ultra-low-latency systems, the dominant inference cost can shift from arithmetic to feature transport latency. In such systems, the order in which features arrive is critical: conventional pipelines buffer the full sample and then process it uniformly, leaving computation idle while salient features may already be available. We address this by structuring both the transmission schedule and the inference computation around feature salience derived from a fully-connected model, rather than input order. All decisions are made at design time; there is no runtime branching or adaptive control, ensuring deterministic, hardware-friendly execution.

\subsection{Aligning Compute Allocation with Information Arrival}
\label{subsec:compute_allocation}

Standard FPGA inference pipelines delay computation until input transfer is complete, giving latency proportional to transfer time. We instead structure inference as a cascade: computation proceeds as soon as a feature subset arrives. Let \(B_{\mathrm{io}}\) be the physical input width and \(B_{\mathrm{s}}\) the quantized-bit budget assigned to one feature stage. A stage therefore occupies \(\lceil B_{\mathrm{s}}/B_{\mathrm{io}}\rceil\) input cycles and contains at most \(\lfloor B_{\mathrm{s}}/\beta\rfloor\) features. Partial feature prefixes are used to produce intermediate predictions, which are incrementally refined as additional subsets arrive, eliminating the need for full-sample buffering.

The cascade is statically parameterized by a per-layer gating vector $\mathbf{g} = (g_0, \dots, g_{L-1})$, where $g_\ell \in \{0,1\}$ indicates whether hidden layer $\ell$ consumes newly arriving features in addition to the previous hidden state. Let $h_\ell$ denote the quantized hidden output of layer $\ell$, and let $c_\ell$ denote the feature subset presented at stage $\ell$ when $g_\ell = 1$. Let $q_{\mathrm{in}}(\cdot)$ denote the learned input quantizer applied independently to each feature subset. Here, $\|$ denotes vector concatenation along the feature dimension.
The resulting layer input is
\begin{equation}
  z_\ell =
  \begin{cases}
    q_{\mathrm{in}}(c_\ell),
      & \ell = 0,\; g_0 = 1, \\
    h_{\ell-1},
      & g_\ell = 0, \\
    \left[ h_{\ell-1} \,\|\, q_{\mathrm{in}}(c_\ell) \right],
      & g_\ell = 1,\; \ell > 0,
  \end{cases}
  \label{eq:cascade_input}
\end{equation}
and the hidden update is $h_\ell = f_\ell(z_\ell)$, where $f_\ell$ is implemented as a sparse LUT network layer.
Consequently, the dataflow and pipeline structure are fully determined \textit{a priori}, enabling deterministic, fixed-latency streaming inference.

The gating vector is a compile-time schedule. Entries with \(g_\ell=1\) are feature stages, which place layer \(\ell\) at a feature-injection boundary: its subset is loaded over \(\lceil B_{\mathrm{s}}/B_{\mathrm{io}}\rceil\) cycles, concatenated with \(h_{\ell-1}\) for \(\ell>0\) and registered, as shown in subsets two and three in Figure~\ref{fig:lut-diagram}. Layers with \(g_\ell=0\) consume only hidden activations and are not separated by registers in our implementation, increasing representational capacity through further nonlinear transformations without additional feature-transfer cost. The resulting dataflow is fixed at synthesis time, with no data-dependent stalls or branches.

\subsection{Input Feature Saliency and Ordering}
\label{subsec:feat_order}

To fully exploit the proposed topology, features must be transmitted sequentially in order of salience so that the most salient inputs are available earliest in the stream and can influence the network for the longest duration. To achieve this, we first train a fully connected model in which each cascade stage observes the complete input vector. The trained dense weights are then used to derive neuron-level top-$k$ input masks by selecting the largest-magnitude weights for each neuron, consistent with the masking procedure in~\cite{assemble}.

Feature salience is computed per feature stage by counting how often each input feature appears in the top-$k$ masks across all layers and neurons for stage $s$ in the full-input surrogate:
\begin{equation}
I_i^{(s)} = \sum_{\ell,n} \mathbf{1}[i \in \mathrm{TopK}^{(s)}_{\ell,n}].
\end{equation}
Here, $\mathbf{1}[\cdot]$ is the indicator function, and $\mathrm{TopK}{\ell,n}^{(s)}$ is the set of input indices selected by the top-$k$ mask of neuron $n$ in layer $\ell$ for stage $s$. Features are then ordered by descending salience and greedily assigned, without overlap, to subsets of $\lfloor B_{\mathrm{s}} / \beta \rfloor$ features each. The resulting permutation is fixed after training and reused for all samples statically at runtime.

This ordering schedules the most discriminative features in earlier load cycles, enabling accurate prediction from partial inputs and refinement as additional features arrive.

\subsection{Per-Feature Learned Quantization}
\label{subsec:encoding}

Under a fixed link budget of $B_{\mathrm{io}}$ bits per cycle, reducing end-to-end latency requires transmitting features in a form that preserves their discriminative utility while minimizing bitwidth. We therefore learn a low-bit quantization scheme that maps each feature $x_i$ to a discrete value prior to transmission. 

Each feature is quantized by learned thresholds ${t_{i,j}}$ and scale $s_i$. The comparisons form an internal thermometer vector, but only its population count is exposed:
\begin{equation}
z_{i,j}=s_i(x_i-t_{i,j}), \quad q_i(x_i)=\sum_j \mathbf{1}[z_{i,j}>0].
\end{equation}
This count is encoded as a compact $\beta$-bit unsigned binary value consumed by CascadeLUT. Thus, on-device preprocessing produces the same binary code through L-LUT quantizers.

Thresholds are parameterized as a base value plus positive increments to ensure strict ordering during training. Because the hard quantizer is non-differentiable, training uses a straight-through estimator with a clipped linear surrogate and a learnable temperature parameter.

At inference, the encoder maps to a bank of comparators implemented as one L-LUT per feature.
This forms an L-LUT quantization layer placed at the input of each feature stage, matching the network representation and preserving the LUT-based computation model. The quantization operates within the same load cycle as feature arrival, so it does not introduce additional pipeline stages, latency or buffering.

\begin{figure}[b]
  \centering
  \includegraphics[width=\linewidth]{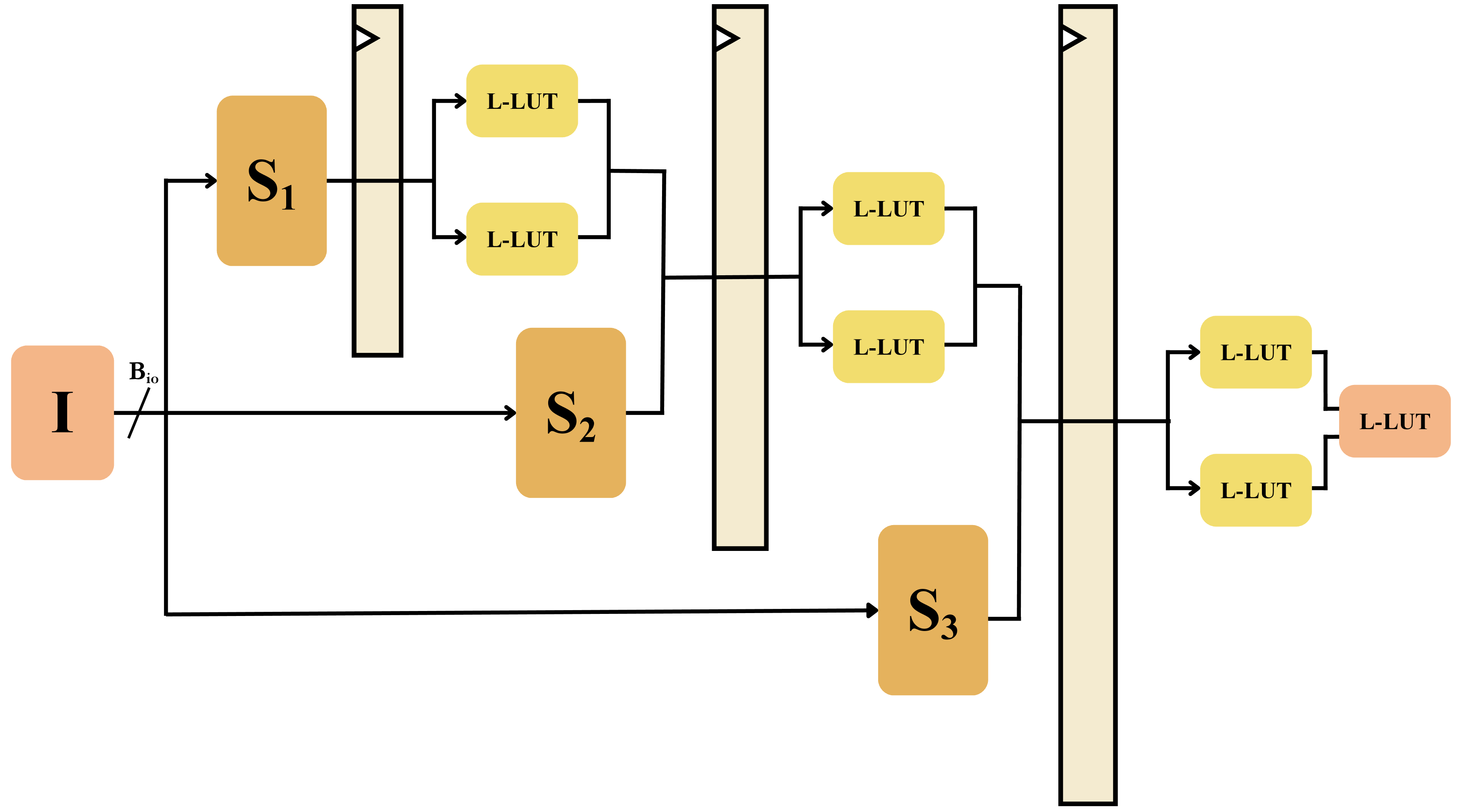}
  \caption{CascadeLUT Topology. Feature subsets \(S_s\) stream into fixed pipeline boundaries, avoiding full-sample buffering.}
  \label{fig:lut-diagram}
\end{figure}

\subsection{Training and Toolflow}

The CascadeLUT pipeline consists of three training phases followed by hardware mapping.

\subsubsection{Feature Ordering}
A fully connected model is trained where each cascade stage observes the full input vector. Feature salience scores are then derived following Section~\ref{subsec:feat_order} to front-load discriminative information under the bandwidth constraint. This provides an inexpensive proxy for feature salience, avoiding costly gradient-based attribution methods.

\subsubsection{Sparse Cascade Training}
Using the derived feature schedule, the cascade network is retrained with the bandwidth-constrained input structure. Sparse connectivity masks are extracted from the fully connected weights using the magnitude-based fan-in selection as in prior LUT-based architectures~\cite{assemble}.

\subsubsection{Final Training}
The resulting sparse cascade network is retrained under the final connectivity constraints using SGD with decoupled weight decay~\cite{decoupled} and warm restarts~\cite{sgd}.

\subsubsection{RTL Generation}
Each neuron is mapped to a lookup-table node by enumerating its learned Boolean function. Pipeline registers are inserted between feature stages to meet timing at the target clock frequency.

\begin{figure}[b]
  \centering
  \includegraphics[width=0.9\linewidth]{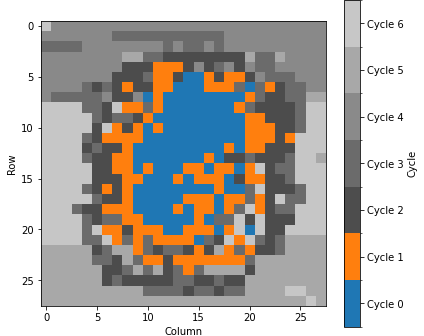}
  \caption{MNIST feature loading order by cycle. Colored features are loaded by the model; gray features are truncated.}
  \label{fig:mnist-sample}
\end{figure}

\section{Results}\label{sec:results}

\subsection{Experimental Setup}

\begin{table*}[t]
\centering
\scriptsize
\setlength{\tabcolsep}{3pt}
\renewcommand{\arraystretch}{1.12}
\caption{CascadeLUT model architectures and training parameters used across our evaluation.}
\label{table:assemble-configs}
\begin{tabular}{@{}lcccccccc@{}}
\toprule
Dataset & $w_\ell$ & $g_\ell$ & $F$ & $\beta$ & $(L,N,S)$ & $B_{\mathrm{s}}$ & II & $N_u/N_{\mathrm{in}}$ \\
\midrule
MNIST &
$[2160,360,2160,360,60,10]$ &
$[1,0,1,0,0,0]$ &
$[6,6,8,6,6,6]$ &
$[1,1,1,1,1,6]$ &
$(2,75,2)$ & 112 & 2 & $224/784$ \\

FashionMNIST &
$[2160,360,2400,1000,2160,360,60,10]$ &
$[1,0,1,1,1,0,0,0]$ &
$[8,6,8,8,8,6,6,6]$ &
$[1,1,1,1,1,1,1,6]$ &
$(2,64,2)$ & 112 & 4  & $448/784$ \\

KWS &
$[1600,400,600,704,176,44,11]$ &
$[1,0,1,1,0,0,0]$ &
$[4,4,4,4,4,4,4]$ &
$[2,2,2,2,2,2,6]$ &
$(2,92,2)$ & 224 & 6  & $336/640$ \\

ToyADMOS/car &
$[500,512,128,32,8,2]$ &
$[1,1,0,0,0,0]$ &
$[4,4,4,4,4,4]$ &
$[2,2,2,2,2,4]$ &
$(2,75,2)$ & 224 & 4  & $224/533$ \\
\bottomrule
\end{tabular}
\end{table*}

Table~\ref{table:assemble-configs} lists all evaluated configurations and should be read with Fig.~\ref{fig:lut-diagram}: \(g_\ell=1\) entries instantiate the orange feature-injection blocks \(S_s\), while \(g_\ell=0\) layers use only hidden activations. \(B_{\mathrm{s}}\) gives each block's bit budget, II the initiation interval and \(N_u/N_{\mathrm{in}}\) the consumed/available original features after truncation. Layer sizes \(w_\ell\) give the number of L-LUTs, while \(F\) and \(\beta\) define fan-in and quantization. The NeuraLUT-Assemble subnetwork dimensions are \((L,N,S)\)~\cite{assemble}.

Evaluation uses MNIST~\cite{mnist}, FashionMNIST~\cite{fashionmnist}, Google Speech Commands keyword spotting (KWS)~\cite{kws} and ToyADMOS anomaly detection~\cite{toyadmos}. KWS is an 11-class audio keyword-spotting task based on Google Speech Commands features, while ToyADMOS/car is a binary acoustic anomaly-detection task for machine monitoring. During training, inputs remain floating-point to optimize the learned input encoders.

For hardware deployment we consider two regimes. In the first, inputs are preprocessed offline and quantized features are streamed directly to the FPGA. In the second, raw \texttt{int8} features are streamed and all preprocessing is performed on-chip using learned threshold L-LUTs, reflecting deployments where sensor data is transmitted directly to the device.

All experiments are conducted on a Xilinx Zynq Z-7045 FPGA, as used in previous works. Designs are synthesized, placed and routed at 200 MHz, with inputs streamed via physical I/O pins under a fixed \(B_{\mathrm{io}}=112\) bits/cycle constraint. Table~\ref{table:assemble-configs} reports \(B_s\), the quantized-bit budget per cascade feature stage; each stage therefore takes $\lceil B_s/B_{\mathrm{io}}\rceil$ input cycles, so \(B_s=224\) denotes two 112-bit cycles rather than a wider link. Reported latency reflects full end-to-end inference, including feature transfer, enabling fair comparison under identical I/O constraints. LUT utilization corresponds to post-implementation usage. Throughput is derived from cycle count and energy per sample from dynamic power over throughput.
Test accuracy is evaluated on the official splits for MNIST/FashionMNIST and the benchmark splits from ULEEN for KWS/ToyADMOS~\cite{uleen}.

\subsection{Bandwidth Optimization Results}
\label{subsec:binary_inputs}
\begin{table*}[t]
\centering
\footnotesize
\caption{Implementation results for CascadeLUT and prior FPGA inference architectures on a Zynq Z-7045. Preprocessed inputs are streamed via I/O pins (112\,bits/cycle). Clock frequency is fixed at 200\,MHz to match published Zynq Z-7045 baselines.  
\textdagger~Data augmentation used. *Model could not be synthesized; the hardware values are approximate. }
\label{tab:dwn_results}
\resizebox{\textwidth}{!}{
\begin{tabular}{llcccccc}
\hline
\textbf{Dataset} &
\textbf{Model} &
\textbf{Test Acc. (\%)} &
\textbf{Latency (ns)} &
\textbf{Throughput (Samples/s)} &
\textbf{Energy (nJ/Sample)} &
\textbf{LUTs (k)} \\
\hline

\multirow{9}{*}{MNIST}
& FINN                & 98.40 & 2440 & 1.56M & 5445 & 83.0 \\
& ULEEN\textsuperscript{\textdagger}    & 98.46 & 940  & 4.05M & 823  & 123.1 \\
& DiffLogicNet (xs)   & 96.87 & 90   & 33.3M & 17.2 & 9.6 \\
& DWN ($n{=}6$; lg)   & \textbf{98.31} & 125  & 25.0M & 19.0 & \textbf{4.6} \\
& DWN ($n{=}6$; lg; +aug)\textsuperscript{\textdagger}
                      & \textbf{98.77} & 135  & 22.2M & 21.6 & \textbf{4.6} \\
& NeuraLUT-Assemble & 97.91 & 40 & 28.6M & 9.4 & 5.1 \\ 
& NeuraLUT-Assemble\textsuperscript{\textdagger} & 98.59 & 40 & 28.6M & 9.5 & 5.1 \\ 
& \textbf{CascadeLUT} & 98.01 & \textbf{10} & \textbf{100.0M} & \textbf{4.3} & 10.3 \\ 
& \textbf{CascadeLUT\textsuperscript{\textdagger}} & 98.59 & \textbf{10} & \textbf{100.0M} & \textbf{4.5} & 10.3 \\ 
\hline

\multirow{5}{*}{FashionMNIST}
& FINN              & 84.36 & 2440 & 1.56M & 5445 & 83.0 \\
& ULEEN             & 87.86 & 940  & 4.05M & 823  & 123.1 \\
& DiffLogicNet      & 87.44 & 270  & 9.52M & 119.8 & 11.4 \\
& DWN ($n{=}6$)     & 89.01 & 250  & 10.0M & 90.9 & \textbf{7.6} \\
& \textbf{CascadeLUT} & \textbf{89.14} & \textbf{20} & \textbf{50.0M} & \textbf{6.6} & 12.2 \\ 
\hline

\multirow{5}{*}{KWS}
& FINN              & 70.60 & 7780 & 0.67M & 5716 & 42.8 \\
& ULEEN             & 70.34 & 390  & 10.0M & 642  & 141.1 \\
& DiffLogicNet*     & 64.18 & 265  & 10.0M & ---  & 20.3 \\
& DWN ($n{=}6$)     & 71.52 & 235  & 10.5M & 42.3 & \textbf{4.8} \\
& \textbf{CascadeLUT} & \textbf{71.55} & \textbf{30} & \textbf{33.3M} & \textbf{10.0} & 20.9 \\ 
\hline

\multirow{5}{*}{ToyADMOS/car}
& FINN              & 86.10 & 3520 & 1.57M & 547  & 14.1 \\
& ULEEN             & 86.33 & 340  & 11.1M & 143  & 29.4 \\
& DiffLogicNet      & 86.66 & 165  & 16.7M & 23.2 & \textbf{3.9} \\
& DWN ($n{=}6$; lg) & \textbf{89.03} & 165  & 16.7M & 45.4 & 6.2 \\
& \textbf{CascadeLUT} & 88.07 & \textbf{20} & \textbf{50.0M} & \textbf{5.6} & 7.3 \\ 
\hline

\end{tabular}
}
\end{table*}

\begin{table*}[t]
\centering
\footnotesize
\renewcommand{\arraystretch}{1.15}
\caption{Implementation results for CascadeLUT and prior work on a Zynq Z-7045 with on-device preprocessing. Inputs are streamed via I/O pins (112 bits/cycle) as \texttt{int8} at 200 MHz clock frequency. $\Delta$LUTs denotes preprocessing overhead.}
\label{tab:in_quant}
\resizebox{\textwidth}{!}{
\begin{tabular}{llccccccc}
\hline
\textbf{Dataset} &
\textbf{Model} &
\textbf{Test Acc. (\%)} &
\textbf{Latency (ns)} &
\textbf{Throughput (Samples/s)} &
\textbf{Energy (nJ/Sample)} &
\textbf{LUTs (k)} &
\textbf{$\Delta$LUTs (k)}\\
\hline

\multirow{2}{*}{MNIST}
& NeuraLUT-Assemble & 97.91 & 260 & 3.92M & 36.7 & \textbf{7.6} & 2.5 \\ 
& \textbf{CascadeLUT} & \textbf{98.01} & \textbf{80} & \textbf{12.5M} & \textbf{20.1} & 10.8 & \textbf{0.5} \\ 
\hline

\multirow{1}{*}{FashionMNIST}
& \textbf{CascadeLUT} & 89.14 & 160 & 6.25M & 36.2 & 13.7 & 1.5 \\ 
\hline

\multirow{1}{*}{KWS}
& \textbf{CascadeLUT} & 71.53 & 120 & 8.33M & 31.9 & 24.1 & 3.2 \\ 
\hline

\multirow{1}{*}{ToyADMOS/car}
& \textbf{CascadeLUT} & 88.06 & 80 & 12.5M & 19.3 & 8.1 & 0.8 \\ 
\hline

\end{tabular}
}
\end{table*}

In this configuration, input samples are quantized offline and streamed to the FPGA as preprocessed feature vectors over the 112-bit I/O interface. Table~\ref{tab:dwn_results} summarizes implementation results alongside prior FPGA inference architectures.

CascadeLUT achieves the lowest inference latency on every task, with speedups of 4.0--12.5\(\times\) over the fastest prior work. 
Latency is determined by feature bitwidth $\beta$ and the number of feature stages, which control how quickly salient features are delivered under fixed bandwidth.
For MNIST and FashionMNIST, binarized inputs allow 112 features per cycle, concentrating salient information early in the stream and enabling very low latency, with FashionMNIST requiring two additional loads of background pixels to maintain accuracy.
On MNIST, CascadeLUT improves over NeuraLUT-Assemble by eliminating non-load pipeline
stages, so latency is determined by feature-transfer cycles. Predictions are formed after only two feature loads, reducing end-to-end latency by 4$\times$.

KWS and ToyADMOS use 2-bit representations, halving features per cycle and requiring more loads to reach comparable coverage, which increases latency. However, the advantage over DWN remains because the cascade produces useful predictions from the first arriving subsets rather than waiting for the full input. This introduces a trade-off: lower-precision features increase features per cycle and accelerate early coverage, while higher-precision features reduce features per cycle and delay sufficient coverage for inference.

Energy per sample scales with latency, as dynamic power is consumed only while the pipeline is active and thus depends on the number of active cycles per sample. CascadeLUT reduces this by minimizing load stages, achieving up to $13.8\times$ energy improvement over DWN across tasks.

CascadeLUT achieves competitive accuracy across all tasks despite operating on partial input subsets at each stage, achieving the highest reported accuracy on FashionMNIST and KWS. This confirms that the learned feature ordering preserves sufficient information for accurate classification even when large portions of the input are delayed or truncated.

The cascade topology incurs an area cost relative to the most compact baselines. Each feature stage requires increased LUT fan-in due to the concatenation of new inputs with propagated activations, and interleaving hidden layers between stages adds depth relative to standard architectures. Consequently, CascadeLUT uses 1.2--4.4\(\times\) the LUTs that the smallest DWN configurations use on each task, with the gap widest on KWS where reduced sparsity limits pruning opportunities. However, it remains substantially smaller than ULEEN, which requires 4--12\(\times\) more LUTs across the tasks. CascadeLUT occupies an intermediate area–latency design point, trading increased area for significantly reduced latency under bandwidth constraints.

\subsection{MNIST Feature Ordering}

CascadeLUT reduces latency by prioritizing the transmission of salient features. Inputs are ordered so that the most class-discriminative features are transferred first, while less relevant features are delayed or omitted entirely. As a result, accurate predictions can be formed without requiring the full input vector to be transferred to the device.

Figure~\ref{fig:mnist-sample} visualizes the learned pixel loading order for MNIST following Section~\ref{subsec:feat_order}. Central pixels, which contain the most discriminative structure, are transmitted first, while peripheral pixels are scheduled later and many background pixels are never loaded. As a result, most discriminative information is captured within the first two cycles—just 224 of 784 pixels—sufficient for accurate prediction and yielding an end-to-end latency of 10 ns at a 200 MHz clock frequency.

\subsection{Input Preprocessing Results}
\label{subsec:on_device_quant}

On-chip quantization uses threshold comparators, mapping each feature to a single L-LUT that jointly implements batch normalization and learned thresholding. Batch normalization reduces to an affine transform, and is followed by comparison so both operations can be folded into LUT logic.

Relative to the offline configuration, latency increases by $8\times$ for MNIST and FashionMNIST (previously binary) and $4\times$ for KWS and ToyADMOS (previously 2-bit), since the compute pipeline remains unchanged and transfer cost scales with input bitwidth. Energy per sample increases because higher-precision inputs require more cycles, extending inference duration rather than increasing switching in the quantization\,stage.

Applying the same per-feature L-LUT quantization to NeuraLUT-Assemble yields similar latency increases. Consequently, CascadeLUT retains its latency advantage by eliminating non-load pipeline stages and reducing load stages.

Area overhead, reported as $\Delta$LUTs, corresponds to the additional LUT resources required to implement on-device input preprocessing. This overhead remains modest and scales with both the number of feature stages and the quantization bitwidth $\beta$, which increase the number and cost of per-feature quantizers.
Across all tasks, overhead is reduced by feature truncation, as quantizers are only instantiated for features used by the cascade. Compared to NeuraLUT-Assemble on MNIST, CascadeLUT reduces on-device quantization resources by $5\times$.

\section{Conclusion and Future Work}
\label{sec:conc}

We presented CascadeLUT, a framework for bandwidth-constrained NN inference. By structuring network topology around the streaming data path, CascadeLUT achieves 4.0--12.5\(\times\) lower latency, 3.0--5.0\(\times\) higher throughput and up to 13.8\(\times\) lower energy per sample than the fastest prior baseline per task while maintaining comparable accuracy. These gains use additional LUTs for lower latency: CascadeLUT uses 1.2--4.4\(\times\) the LUTs of the smallest DWN baseline because feature injection increases fan-in and inter-stage logic. 
Future work will explore extending the cascade framework to additional architectures and deployment scenarios, including design-time feature ordering adapted to task-specific input distributions and integration with hardware-aware training techniques.

\IEEEtriggeratref{12}
\bibliographystyle{IEEEtran}
\bibliography{bibliography}

@inproceedings{finn,
	author		= "{Y. Umuroglu \emph{et al.}}",
	title		= "{FINN: A Framework for Fast, Scalable Binarized Neural Network Inference}",
	booktitle	= "{Proceedings of the 2017 ACM/SIGDA International Symposium on Field-Programmable Gate Arrays}",
	series  	= "{FPGA '17}",
	pages		= "65–74",
	address 	= "{Monterey, CA, USA}",
	publisher	= "{Association for Computing Machinery}",
	year		= 2017,
	month		= feb,
	note		= "doi: 10.1145/3020078.3021744"
}

@ARTICLE{efficientdnn,
  author={Sze, Vivienne and Chen, Yu-Hsin and Yang, Tien-Ju and Emer, Joel S.},
  journal={Proceedings of the IEEE}, 
  title={Efficient Processing of Deep Neural Networks: A Tutorial and Survey}, 
  year={2017},
  volume={105},
  number={12},
  pages={2295-2329},
  doi={10.1109/JPROC.2017.2761740}}

@article{hls4ml,
    author = "Duarte, Javier and others",
    title = "{Fast inference of deep neural networks in FPGAs for particle physics}",
    eprint = "1804.06913",
    archivePrefix = "arXiv",
    primaryClass = "physics.ins-det",
    reportNumber = "FERMILAB-PUB-18-089-E",
    doi = "10.1088/1748-0221/13/07/P07027",
    journal = "JINST",
    volume = "13",
    number = "07",
    pages = "P07027",
    year = "2018"
}

@article{hls4ml-bin,
    author = "Ngadiuba, Jennifer and others",
    title = "{Compressing deep neural networks on FPGAs to binary and ternary precision with HLS4ML}",
    eprint = "2003.06308",
    archivePrefix = "arXiv",
    primaryClass = "cs.LG",
    reportNumber = "FERMILAB-PUB-20-167-PPD-SCD",
    doi = "10.1088/2632-2153/aba042",
    journal = "Mach. Learn. Sci. Tech.",
    volume = "2",
    pages = "015001",
    year = "2021"
}

@inproceedings{logicnets,
	author		= "Umuroglu, Yaman and Akhauri, Yash and Fraser, Nicholas J. and Blott, Michaela",
	title		= "{LogicNets: Co-Designed Neural Networks and Circuits for Extreme-Throughput Applications}",
	booktitle	= "{2020 30th International Conference on Field-Programmable Logic and Applications (FPL)}",
	pages		= "291-297",
	address 	= "{Gothenburg, Sweden}",
	publisher	= "{IEEE Computer Society}",
	year		= 2020,
	month		= sep,
	note		= "doi: 10.1109/FPL50879.2020.00055"
}

@inproceedings{poly,
	author		= "Andronic, Marta and Constantinides, George A.",
	title		= "{PolyLUT: Learning Piecewise Polynomials for Ultra-Low Latency FPGA LUT-based Inference}",
	booktitle	= "{2023 International Conference on Field Programmable Technology (ICFPT)}",
	pages		= "60-68",
	address 	= "Yokohama, Japan",
	publisher	= "IEEE",
	year		= 2023,
	month		= dec,
	note		= "doi: 10.1109/ICFPT59805.2023.00012"
}

@article{polylut2,
  author  = {Marta Andronic and Jiawen Li and George A. Constantinides},
  title   = {{PolyLUT}: Ultra-Low Latency Polynomial Inference with Hardware-Aware Structured Pruning},
  journal = {{IEEE} Transactions on Computers},
  volume  = {74},
  number  = {9},
  pages   = {3181--3194},
  year    = {2025},
  doi     = {10.1109/TC.2025.3586311}
}

@inproceedings{NeuraLUT,
	author		= "Andronic, Marta and Constantinides, George A.",
	title		= "{NeuraLUT: Hiding Neural Network Density in Boolean Synthesizable Functions}",
	booktitle	= "{2024 34th International Conference on Field-Programmable Logic and Applications (FPL)}",
	pages		= "140-148",
	publisher	= "IEEE",
	year		= 2024,
	note		= "doi: 10.1109/FPL64840.2024.00028"
}

@inproceedings{assemble,
	author	= "Andronic, Marta and Constantinides, George A.",
	title		= "{NeuraLUT-Assemble: Hardware-Aware Assembling of Sub-Neural Networks for Efficient LUT Inference}",
	booktitle	= "{2025 IEEE 33rd Annual International Symposium on Field-Programmable Custom Computing Machines (FCCM)}",
	pages		= "208-216",
	publisher	= "IEEE",
	year		= 2025,
	note		= "doi: 10.1109/FCCM62733.2025.00077"
}

@inproceedings{dwn,
  author    = {Alan Tendler Leibel Bacellar and Zachary Susskind and Mauricio Breternitz Jr. and Eugene John and Lizy Kurian John and Priscila Machado Vieira Lima and Felipe M. G. Fran{\c{c}}a},
  title     = {Differentiable Weightless Neural Networks},
  booktitle = {Proceedings of the 41st International Conference on Machine Learning},
  series    = {Proceedings of Machine Learning Research},
  volume    = {235},
  pages     = {2277--2295},
  year      = {2024},
  publisher = {PMLR}
}

@article{uleen,
author = {Susskind, Zachary and Arora, Aman and Miranda, Igor D. S. and Bacellar, Alan T. L. and Villon, Luis A. Q. and Katopodis, Rafael F. and de Ara\'{u}jo, Leandro S. and Dutra, Diego L. C. and Lima, Priscila M. V. and Fran\c{c}a, Felipe M. G. and Breternitz Jr., Mauricio and John, Lizy K.},
title = {{ULEEN}: A Novel Architecture for Ultra-low-energy Edge Neural Networks},
year = {2023},
issue_date = {December 2023},
publisher = {Association for Computing Machinery},
address = {New York, NY, USA},
volume = {20},
number = {4},
issn = {1544-3566},
url = {https://doi.org/10.1145/3629522},
doi = {10.1145/3629522},
journal = {ACM Trans. Archit. Code Optim.},
month = dec,
articleno = {61},
numpages = {24}
}

@inproceedings{difflogic,
  author    = {Felix Petersen and Christian Borgelt and Hilde Kuehne and Oliver Deussen},
  title     = {Deep Differentiable Logic Gate Networks},
  booktitle = {Advances in Neural Information Processing Systems},
  volume    = {35},
  pages     = {2006--2018},
  year      = {2022}
}

@inproceedings{reducedlut,
author = {Cassidy, Oliver and Andronic, Marta and Coward, Samuel and Constantinides, George A.},
title = "{ReducedLUT: Table Decomposition with ``Don't Care'' Conditions}",
year = {2025},
isbn = {9798400713965},
publisher = {Association for Computing Machinery},
address = {New York, NY, USA},
url = {https://doi.org/10.1145/3706628.3708823},
doi = {10.1145/3706628.3708823},
booktitle = {Proceedings of the 2025 ACM/SIGDA International Symposium on Field Programmable Gate Arrays},
pages = {36–42},
numpages = {7},
location = {Monterey, CA, USA},
series = {FPGA '25}
}

@inproceedings{compressedlut,
author = {Khataei, Alireza and Bazargan, Kia},
title = {{CompressedLUT}: An Open Source Tool for Lossless Compression of Lookup Tables for Function Evaluation and Beyond},
year = {2024},
isbn = {9798400704185},
publisher = {Association for Computing Machinery},
address = {New York, NY, USA},
url = {https://doi.org/10.1145/3626202.3637575},
doi = {10.1145/3626202.3637575},
booktitle = {Proceedings of the 2024 ACM/SIGDA International Symposium on Field Programmable Gate Arrays},
pages = {2–11},
numpages = {10},
location = {Monterey, CA, USA},
series = {FPGA '24}
}

@inproceedings {yukio,
  author    = {Yukio Miyasaka and Alan Mishchenko and John Wawrzynek and Nicholas J. Fraser},
  title     = {Synthesis of {LUT} Networks for Random-Looking Dense Functions with Don't Cares---Towards Efficient {FPGA} Implementation of {DNN}},
  booktitle = {2024 IEEE 32nd Annual International Symposium on Field-Programmable Custom Computing Machines},
  series    = {FCCM},
  pages     = {126--132},
  month     = may,
  year      = {2024},
  doi       = {10.1109/FCCM60383.2024.00023}
}

@inproceedings{decoupled,
title={Decoupled Weight Decay Regularization},
author={Ilya Loshchilov and Frank Hutter},
booktitle={International Conference on Learning Representations},
year={2019},
url={https://openreview.net/forum?id=Bkg6RiCqY7},
}

@inproceedings{sgd,
	author		= "Ilya Loshchilov and Frank Hutter",
	title		= "{SGDR: Stochastic Gradient Descent with Warm Restarts}",
	booktitle	= "{5th International Conference on Learning Representations}",
	address 	= "{Toulon, France}",
	year		= 2017,
	month		= apr,
	url         = "https://openreview.net/forum?id=Skq89Scxx"
}

@misc{mnist,
    author    = "Yan LeCun and Corrina Cortes and Christopher J.C. Burges",
    title     = "{The MNIST database of handwritten digits}",
    howpublished = {[Online]. Available: \url{http://yann.lecun.com/exdb/mnist/}},
      year         = {1998},
      note         = {Accessed: 2026-01-20}
}

@misc{fashionmnist,
  author        = {Han Xiao and Kashif Rasul and Roland Vollgraf},
  title         = {{Fashion-MNIST}: A Novel Image Dataset for Benchmarking Machine Learning Algorithms},
  year          = {2017},
  eprint        = {1708.07747},
  archivePrefix = {arXiv},
  primaryClass  = {cs.LG},
  doi           = {10.48550/arXiv.1708.07747}
}

@misc{kws,
  author        = {Pete Warden},
  title         = {Speech Commands: A Dataset for Limited-Vocabulary Speech Recognition},
  year          = {2018},
  eprint        = {1804.03209},
  archivePrefix = {arXiv},
  primaryClass  = {cs.CL},
  doi           = {10.48550/arXiv.1804.03209}
}

@inproceedings{toyadmos,
  author    = {Yuma Koizumi and Shoichiro Saito and Hisashi Uematsu and Noboru Harada and Keisuke Imoto},
  title     = {{ToyADMOS}: A Dataset of Miniature-Machine Operating Sounds for Anomalous Sound Detection},
  booktitle = {2019 IEEE Workshop on Applications of Signal Processing to Audio and Acoustics},
  series    = {WASPAA},
  pages     = {313--317},
  year      = {2019},
  doi       = {10.1109/WASPAA.2019.8937164}
}

\end{document}